\documentclass[pra,twocolumn,showpacs,amsmath,amssymb,superscriptaddress,unsortedaddress,floatfix]{revtex4}

\usepackage{graphicx}% Include figure files
\usepackage{dcolumn}% Align table columns on decimal point
\usepackage{bm}% bold math
\usepackage{color}
\usepackage{bbm}
\usepackage{amsmath}
\newcommand{\Tr}{\textrm{Tr}}

\def\duzomniejsze{<\kern-.7mm<}
\def\duzowieksze{>\kern-.7mm>}

\def\textbf#1{{\bf #1}}
\def\beq{\begin{equation}}
\def\eeq{\end{equation}}
\def\be{\begin{equation}}
\def\ee{\end{equation}}
\def\ben{\begin{eqnarray}}
\def\een{\end{eqnarray}}
\def\beqa{\begin{eqnarray}}
\def\eeqa{\end{eqnarray}}
\def\eea{\end{array}}
\def\bea{\begin{array}}
\newcommand{\bei}{\begin{itemize}}
\newcommand{\eei}{\end{itemize}}
\newcommand{\bee}{\begin{enumerate}}
\newcommand{\eee}{\end{enumerate}}

\begin{document}

%\preprint{}

\title{The decay of quantum correlations between quantum dot spin qubits
and the characteristics of its magnetic field dependence}% Force line breaks with \\

\author{Pawe{\l} Mazurek}
\affiliation{Institute for Theoretical Physics and Astrophysics,
University of Gda{\'n}sk, 80-952 Gda{\'n}sk, Poland}
\affiliation{National Quantum Information Centre of Gda{\'n}sk, 81-824 Sopot, Poland}

\author{Katarzyna Roszak}
\affiliation{Institute of Physics, Wroc{\l}aw University of Technology,
50-370 Wroc{\l}aw, Poland}

\author{Pawe{\l} Horodecki}
\affiliation{Faculty of Applied Physics and Mathematics, Gda{\'n}sk University of Technology, 
80-952 Gda{\'n}sk, Poland}
\affiliation{National Quantum Information Centre of Gda{\'n}sk, 81-824 Sopot, Poland}

\date{\today}% It is always \today, today,
             %  but any date may be explicitly specified

\begin{abstract}
We address the question of the role of quantum correlations beyond entanglement 
in context of quantum magnetometry. To this end,
we study the evolution of the quantum discord, measured by the rescaled discord,
of two electron-spin qubits interacting with an environment of nuclear spins
via the hyperfine interaction.
We have found that
depending on the initial state the evolution can or cannot display indifferentiability points
in its time-evolution (due to the energy conservation law), as well as 
non-trivial dependence on inter-qubit phase.
Furthermore, we show that for initial Bell states, quantum correlations
display a strong magnetic-field sensitivity which can be utilized for decoherence-driven
measurements of the external magnetic field. The potential discord-based measurement 
is sensitive to a wider range of magnetic field values than the entanglement-based
measurement. In principle, entanglement is not a necessary resource for 
reliable decoherence-driven measurement, while the presence of quantum correlations beyond
entanglement is.
\end{abstract}

%\pacs{03.67.Lx, 42.50.Dv}% PACS, the Physics and Astronomy
                             % Classification Scheme.
%\keywords{Suggested keywords}%Use showkeys class option if keyword
                              %display desired
\maketitle

\section{Introduction}

Quantum entanglement represents the correlations that cannot be explained in any classical terms
(see \cite{horodecki2009} and references therein). In particular the corresponding entangled state cannot 
be composed using the two ingredients: a product of quantum states and classical randomness shared by two or more observers. 
However there is another type quantum correlations that go beyond entanglement
\cite{ollivier2001,henderson2001,oppenheim02} (for review see \cite{modi12}). 
They are represented by the separable states ie. the ones that 
can be reproduced with help of the two ingredients mentioned above, however 
they reflect {\it noncommutativity} of quantum physics since 
the product states in the classical probabilistic mixture are eigenvectors of more than 
one mutually noncommuting observable.
This noncommutativity is responsible for the fact that quantum correlations beyond entanglement, 
although much weaker than entanglement itself, may outperform  classical resources 
in some quantum information tasks, e. g. the Knill-Laflamme scheme \cite{knill98,datta08}
and the probabilistic solution of the Deutsch-Jozsa problem \cite{biham04}.

Recently quantum correlations beyond entanglement have been utilised in
other communication tasks: retrieving classial information encoded in
quantum correlations \cite{gu12}, remote state preparation \cite{tekst}, and some
protocols of state discrimination \cite{zhang13}.
Quite remarkably is is known that the total set of quantumly correlated states 
(ie. entangled states and quantumly correlated separable states) dominates 
in the set of all states in the sense that a randomly picked state must belong to it 
\cite{ferraro2010} which means that classical states represent a rare objects in set of 
all quantum states.

The most popular measures of quantum correlations beyond entanglement 
are the quantum discord \cite{ollivier2001,henderson2001} and the quantum deficit 
\cite{oppenheim02}. Recently gometric versions of the quantum discord 
(measuring the smallest distance between a given state and the set of zero-discord states)
atracted much attention because of its computable character 
\cite{dakic2010,luo10,nakano13,paula13,spehner13,spehner14},
and methods to compute some of these measures for arbitrary two-qubit density matrices
have been developed \cite{dakic2010,miranowicz2012,tufarelli13}.
While it must be stressed that geometric objects cannot be treated as measures 
exactly (see \cite{piani12}) they serve as a reasonable indicators of quantum correlations 
and have been
improved recently by an extra rescaling procedure \cite{tufarelli13}.

In the context of some positive observations of the usefulness of quantum correlation beyond entanglement 
in quantum information
being made, it is natural to ask whether their presence may be useful 
in some tasks dedicated to specific physical models. 

Systems of electron spins, each confined in a semiconductor quantum dot (QD)
are appealing in the context of the study of inter-qubit quantum correlations. Firstly,
because of the naturally occurring qubit of electron-spin-up (parallel to the applied
magnetic field) and electron-spin-down (anti parallel to the magnetic field)
which is long lived in this solid state scenario.
Secondly, because of the high level of experimental state-of-the-art, which allows
for a wide range of initializable states and measurement 
possibilities \cite{elzerman04,hanson05,petta05,laird2010,lai2011,barthel12,medford12}.
A flagship example of the experimental possibilities is the recently experimentally
demonstrated quantum state tomography performed on two singlet-triplet qubits
(four electrons in four quantum dots) \cite{shulman2012}.
The main decoherence mechanism for electron spins in QDs
is the hyperfine interaction with the spins of the nuclei of the surrounding atoms
(see Refs~[\onlinecite{coish2009,coish2010,cywinski2011}] for review).
This interaction leads to pure dephasing
at moderately high magnetic fields, but at lower magnetic fields
a more involved decoherence process is seen which leads to a redistribution of the electron
spin occupations. The decoherence processes typically occur on nanosecond time-scales.

One of the interesting questions is to ask about the dependence of the behaviour 
of quantum correlations beyond entanglement in electron spin systems 
with respect to an external tunable parameter. This question may lead to the variant 
of quantum magnetometry \cite{budker,wasilewski,maze} with much weaker state resource then quantum entanglement. 
To this aim in the present paper we study the decay of quantum correlations, quantified by the 
rescaled discord \cite{tufarelli13},
between two electron spins, each confined in a separate QD and interacting with
separate nuclear spin environments,
for two classes of pure initial states.
Firstly, the Bell states are studied, the evolution of which preserves Bell-diagonal form
(up to local unitary oscillations)
in the QD scenario \cite{mazurek2013}, which significantly simplifies the study of the discord 
and provides a convenient starting point for the study of magnetic field dependence
of the characteristics of the discord evolution. This magnetic field dependence
is rather involved in the investigated scenario, and it turns out that the more general
quantum correlations described by the discord provide a means to examine the external magnetic field
via two-qubit decoherence
in a wider value range than entanglement based schemes.
Furthermore, Bell state evolution cannot display points of indifferentiability \cite{maziero09,mazzola2010}
which are typical for the decay of the discord for Bell-diagonal states,
because of energy conservation.
We also show that all results pertaining to magnetic field sensing
can in principle be reproduced using separable non-zero-discord states, such as 
appropriately chosen Werner states. 
Furthermore, we study the discord decay for pure initial states for which all density matrix elements
are finite. The discord evolution found is more complex and exhibits both,
points of indifferentiability, and a non-trivial dependence of the decay curves 
on inter-qubit phase factors; these features are not observed for initial Bell states.

The article is organized as follows. The system, the Hamiltonian, and the method of dealing
with the interaction with the environment are described in Sec. \ref{sec_int}.
Sec. \ref{sec_disc} introduces
the rescaled discord, which is the measure used to quantify quantum correlations
in this paper. In Sec. \ref{sec_bell}, the results describing initial Bell states
and initial mixed, Bell-diagonal states are presented. Here, the first subsection
deals with Bell state evolution only, while the second subsection describes the usefulness
of such states for magnetic field measurements, and renounces the necessity of inter-dot
entanglement for such measurements. The third subsection deals with the possibility
of non-differentiable discord evolutions. Sec. \ref{sec_non} describes the evolution
of the rescaled discord for other pure initial states and Sec. \ref{sec_conc}
concludes the paper.
 
\section{Electron spin in a quantum dot system \label{sec_int}}
The system under study consists of two electron spins confined in two, well separated
lateral GaAs QDs. 
Each electron spin constitutes a qubit, with its spin-up (spin parallel to the applied 
magnetic
field) and spin-down (spin anti-parallel to the applied magnetic field) components
indicated as the $|0\rangle$ and $|1\rangle$ qubit states.
We take into account the most common decoherence mechanism for such systems, namely
the hyperfine interaction of each electron spin with the spins of the nuclei of the
surrounding atoms.

Since we assume that the qubits are well separated (there is no inter-qubit interaction
and no overlap between the qubit environments), the Hamiltonian of the whole system 
is of the form
$H=H_{1}\otimes\mathbb{I}_2+\mathbb{I}_1\otimes H_{2}$,
where $H_i$, $i=1,2$ distinguishes between the dots, are single QD Hamiltonians,
and two-qubit evolution may be inferred from the evolutions of a single subsystem.
The single qubit Hamiltonians are given by
\begin{eqnarray}\label{hamiltonian2}
H_{i}&=&-g\mu_B \hat{S}_i^z B
+\sum_{k} A_{k,i}\hat{S}_i^z \hat{I}_{k,i}^z\\
\nonumber
&&+\frac{1}{2}\sum_{k} A_{k,i}\left(\hat{S}_i^+ \hat{I}_{k,i}^-
+\hat{S}_i^- \hat{I}_{k,i}^+\right),
\end{eqnarray}
where the magnetic field B is applied in the $z$ direction. 
The first term in the Hamiltonian (\ref{hamiltonian2}) describes the electron Zeeman splitting,
where $g$ is the effective electron g-factor, $\mu_B$ is the Bohr mangeton,
$\hat{S}_i^z$ is the electron spin component parallel to the magnetic field,
and $B$ is the magnetic field. The remaining two terms describe the hyperfine
interaction, with $\hat{\mathbf{I}}_{k,i}$ denoting the spin operators of nuclei $k$
in dot $i$ and $\hat{\mathbf{S}}_{i}$ denoting the electron spin operators
in dot $i$. The hyperfine interaction is separated into the term with spin operators
parallel to the magnetic field, which is responsible for pure dephasing processes,
and the so called ``flip-flop'' term, which describes possible nuclear-spin-mediated transitions
of the electron spin. This is described with the rising and lowering operators of both
the nuclear and the electron spins, $\hat{I}_{k,i}^{\pm}=\hat{I}_{k,i}^x\pm i \hat{I}_{k,i}^y$
and $\hat{S}_{i}^{\pm}=\hat{S}_{i}^x\pm i \hat{S}_{i}^y$.
The hyperfine coupling constants depend on the species of the nuclei which form each quantum
dot, as well as on the location of each nucleus with respect to the electron wave function, 
$A_{k,i}=A_{k,i}^0|\Psi_i(\mathbf{r}_{k,i})|^2$, where $A_{k,i}^0$ are the coupling
constants of a given nuclear species found at site $k$ in dot $i$ 
(see Ref.~[\onlinecite{mazurek2013}] for details)
and $\Psi_i(\mathbf{r})$ is the wave function of the electron confined in dot $i$.

In general, finding the QD evolution described by the Hamiltonian (\ref{hamiltonian2})
is an involved task, due to the inapplicability of any 
perturbative approach,
because the interaction can be regarded as a small perturbation with respect to the electron Zeeman splitting
only at moderately high magnetic fields \cite{cywinski2011,coish2010,barnes2012}.
The problem simplifies substantially when 
the initial state of the whole system is a product 
of the double QD state and the states of both nuclear reservoirs,
$\sigma(0)=\rho_{DQD}(0)\otimes R_{1}(0)\otimes R_{2}(0)$,
and the initial states of the nuclear baths
are described by infinite-temperature, fully mixed density matrices.
The nuclear baths are well described by infinite-temperature density matrices 
when the nuclear Zeeman energies are very small with respect 
to the thermal energy $k_B T$ \cite{abragam1983,merkulov2002, barnes2011}. Although typical temperatures at which
spin-in-QD experiments are performed are sub-Kelvin, the Zeeman energy splitting
for gallium and arsenate are of the order of $0.1$ neV per Tesla of magnetic field
and the condition is met for the whole range of magnetic fields.
In this case, the unitary coupling model, for which the Hamiltonian (\ref{hamiltonian2})
can be diagonalized exactly, can be applied on short time scales. In this approximation,
all coupling constants are assumed to be the same
and equal to
$\alpha_i=A_i/N_i$, where $A_i=\sum_k A_{k,i}$ and $N_i$ is the number nuclei in dot $i$.
The upper limit on short-time-scale behaviour is approximated by $N_i/A_i$ \cite{cywinski2009},
which turns out to be sufficient for the study of the evolution of the quantum discord.

In the following, parameters
corresponding to two identical lateral GaAs QDs will be used.
All isotopes naturally found in GaAs
carry spin $I=3/2$ and the average hyperfine coupling constant
for this material is $A_i=83$ $\mu$eV \cite{liu2007,cywinski2009,mazurek2013}.
The number of nuclei considered within each dot  
is $N_i = 1.5\cdot 10^6$. Hence, the limit of short-times, when the unitary coupling
model can be safely used is $1.2\cdot 10^4$ ns.

The resulting single QD evolutions depend strongly on the magnetic field.
In the high magnetic field regime, where the condition
$g\mu_B B \gg A$ is met and which corresponds to magnetic fields 
greater than about $3$ T for the parameters used,
the ``flip-flop'' terms are completely negligible.
The resulting evolution is of pure dephasing character and the decay
of the single spin coherence is proportional
to $\exp(-t^2/T^{*2}_2)$, 
with a characteristic constant 
$T^*_2=\sqrt{\frac{6}{I(I+1)}}\sqrt{N}/A$ (as predicted in Ref.~\cite{merkulov2002}). 
$\sqrt{N}/A\approx 10$ ns according to the parameters used
and the $T^*_2=12.36$ ns extracted from the calculation
corroborates this.
The magnetic field dependence in this regime is limited to the frequency of the
unitary oscillations of the electron spin, which are irrelevant with respect
to the quantification of quantum correlations present in a two-qubit system.
At lower magnetic fields, the ``flip-flop'' terms lead to oscillations of 
the QD occupations, which accompany the dephasing process. The amplitude of these 
oscillations is damped with growing magnetic field, while their frequency increases.
For more details see the Supplementary Material of Ref.~[\onlinecite{mazurek2013}]. It is worth mentioning that the mathematical model of the interaction has been recently used to model magnetic sensing carried out by some chemical systems
in biology (see \cite{tiersch13} and references therein).

\section{rescaled discord \label{sec_disc}}

The quantum geometric discord \cite{dakic2010} has stirred up a lot of controversy recently,
being the first quantum discord measure for which straight-forward
formulas (such, that do not involve minimization) 
for the calculations of its lower \cite{dakic2010} and \cite{miranowicz2012} upper bounds
given any two-qubit density matrix
have been found, while being susceptible to increases under local (single-qubit) 
non-unitary evolution, and hence, being an unreliable quantum 
discord measure \cite{piani12}.
The quantum geometric discord is defined as the Hilbert-Schmidt distance between
a given state and the nearest zero-discord state. 
The nature of the problem is related to the use of the Hilbert-Schmidt distance, because this 
particular distance measure is sensitive to the global purity of the studied state.
On the other hand, 
it is the properties of the Hilbert-Schmidt distance that allow for the simple calculation
of the geometric discord for any two-qubit state.

A solution of this problem has been proposed in Ref.~[\onlinecite{tufarelli13}].
It turns out that to diminish the sensitivity of the Hilbert-Schmidt distance
to the purity of the states, it suffices to normalize each state by its Hilbert-Schmidt
norm, namely to define a distance between two states $\rho_1$ and $\rho_2$ as
\begin{equation}
\label{dt}
d_T(\rho_1,\rho_2)=\left|\left|\frac{\rho_1}{||\rho_1||}-\frac{\rho_2}{||\rho_2||}\right|\right|,
\end{equation}
where $||\cdot||$ is the Hilbert-Schmidt norm.
The rescaled discord is then defined as the distance between
a given state and the nearest zero-discord state, using the distance measure (\ref{dt})
and for a two-qubit state it is found to be
\begin{equation}
\label{rd}
D(\rho)=\frac{1}{2}\left( 1-\frac{\sqrt{3}}{2}\right)\left[
1-\sqrt{1-\frac{D_S(\rho)}{2\Tr\rho^2}}
\right].
\end{equation}
Here, $D_S(\rho)$ denotes the geometric discord and $\Tr\rho^2$
is the purity of the studied state.

Hence, there is a straightforward relation between the rescaled discord
and the geometric discord, for which methods of calculation are available
for two qubit states.
The lower bound on the geometric discord
is given by \cite{dakic2010}
\begin{equation}
\label{lower}
D_S'=\frac{1}{4}\max\left(
\Tr[K_x]-k_x,\Tr[K_y]-k_y
\right),
\end{equation}
where $k_x$ is the maximum eigenvalue of the matrix $K_x=|x\rangle\langle x|+TT^T$
and $k_y$ is the maximum eigenvalue of the matrix $K_y=|y\rangle\langle y|+T^T T$.
Here, $|x\rangle$ and $|y\rangle$
denote local Bloch vectors with components $x_i=\Tr[\rho_{AB}(\sigma_i\otimes\mathbb{I})]$
and $y_i=\Tr[\rho_{AB}(\mathbb{I}\otimes\sigma_i)]$, and the elements of the correlation
matrix $T$ are given by $T_{i,j}=\Tr[\rho_{AB}(\sigma_i\otimes\sigma_j)]$ (stemming from the
standard Bloch representation of a two-qubit density matrix $\rho_{AB}$).
The upper bound is given by \cite{miranowicz2012}
\begin{eqnarray}
\label{upper}
D_S''&=&\frac{1}{4}\min\left(
\Tr[K_x]-k_x+\Tr[L_y]-l_y,\right.\\
\nonumber
&&
\left.\Tr[K_y]-k_y+\Tr[L_x]-l_x
\right),
\end{eqnarray}
where $l_x$ and $l_y$ are the maximal eigenvalues of the matrices 
$L_x=|x\rangle\langle x|+T|\hat{k}_y\rangle\langle \hat{k}_y|T^T$ and
$L_y=|y\rangle\langle y|+T^T|\hat{k}_x\rangle\langle \hat{k}_x|T$, respectively,
while $|\hat{k}_x\rangle$ and $|\hat{k}_y\rangle$ are the normalized eigenvectors
corresponding to the eigenvalue $k_x$ of matrix $K_x$ and $k_y$ of matrix $K_y$.
The final step in acquiring the upper and lower bounds on the rescaled
discord  is inserting the geometric discord values into Eq. (\ref{rd}).

The upper and lower bounds of the rescaled discord often coincide, similarly as in the case of the 
geometric discord \cite{roszak13}, yielding its true value.
This is specifically the case for pure states, and
it is straightforward to show that the geometric discord is equal to $1/2$ for all
maximally entangled two-qubit states \cite{blanchard01},
\begin{equation}
\label{ini}
|\psi\rangle =
\sqrt{a}|00\rangle +\sqrt{b}e^{i\alpha}|10\rangle 
+\sqrt{b}e^{i\beta}|01\rangle -\sqrt{a}e^{i(\alpha+\beta)}|11\rangle),
\end{equation}
which are all rotations of Bell diagonal states by tensor products of unitary operations \cite{horodecki96b}. The upper limit is also equal to the lower limit of the rescaled discord for all
X-states (and their subclass, the
Bell diagonal states), and for states with vanishing
local Bloch vectors, $|x\rangle=|y\rangle=0$ \cite{miranowicz2012}.

\section{Bell-diagonal states \label{sec_bell}}

\subsection{Bell states}

Bell states are natural fully entangled states to be studied in a double
spin-in-a-QD system (as in many other realistic scenarios), 
since they are initialized more easily than other
entangled states. The singlet state is distinguished for single spin qubits in
lateral QDs, since its preparation
and the measurement of its fidelity in a double QD has already been demonstrated experimentally
in 2005 \cite{petta05}. The studied evolution does not differentiate between the
Bell states, with the exception of unitary single dot evolutions
which are irrelevant in the study of quantum correlations and have been omitted here.
The evolution preserves the Bell diagonal form (apart from the aforementioned
local unitary oscillations),
as has been shown in Ref.~[\onlinecite{mazurek2013}],
hence the density matrix of the two-qubit system is of the form
\begin{equation}
\label{singlet}
\rho_{DQD}(t)=
\begin{pmatrix}
  \frac{1}{2}-a(t) & 0 & 0 & 0  \\
0& a(t)&b(t) &0 \\
0& b^*(t)&a(t) &0 \\
0&0 &0 &   \frac{1}{2}-a(t)
 \end{pmatrix}
\end{equation}
at any given time. The initial conditions for any Bell state are
$a(t)=1/2$ and $b(t)=\pm 1/2$, with the basis states in the density matrix
(\ref{singlet}) arranged in the order $|\uparrow\uparrow\rangle$, 
$|\uparrow\downarrow\rangle$, $|\downarrow\uparrow\rangle$, 
and $|\downarrow\downarrow\rangle$ 
for $|\Psi^{\pm}\rangle=1/\sqrt{2}(|01\rangle\pm |10\rangle)$ initial states
and in the order $|\uparrow\downarrow\rangle$, $|\uparrow\uparrow\rangle$, 
$|\downarrow\downarrow\rangle$, 
and $|\downarrow\uparrow\rangle$ for $|\Phi^{\pm}\rangle=1/\sqrt{2}(|00\rangle\pm |11\rangle)$ initial states.

As mentioned in the previous section, the lower and upper bounds on the rescaled
discord $D$ coincide for any Bell diagonal state, hence, they coincide
throughout the hyperfine-interaction induced evolution of any initial 
Bell state.
Furthermore, analytical formulas for the values of $D_S$
can be found in a straightforward manner for this type of evolution,
which can then be extended using Eq. (\ref{rd}) to yield the value of $D$.
Indeed the formula for the geometric quantum discord reads
\begin{small}
\begin{equation}
 D_S(\rho_{DQD}(t)) = \left\{\begin{array}{cc} 2|b(t)|^{2} \mbox { for } g(t)\leq 1,  \\
\left[\frac{1}{2}-2a(t)\right]^{2}+|b(t)|^{2} \mbox { for }g(t)\geq 1, \end{array}\right. 
\end{equation}
\end{small}
where 
\begin{equation}
\label{g}
g(t)=\frac{2|b(t)|}{|1-4a(t)|}.
\end{equation}
The purity necessary to find the rescaled discord
is equal to 
\begin{equation}
P(\rho_{DQD}(t))=2\left[ (\frac{1}{2}-a(t))^2+a^2(t)\right]+2|b(t)|^{2}.
\end{equation}
Note, that rescaling $D_{S}$, although it affects the curves of the time-evolution,
does not change the transition point between the two regimes of the discord decay.

\begin{figure}
\includegraphics[scale=0.7]{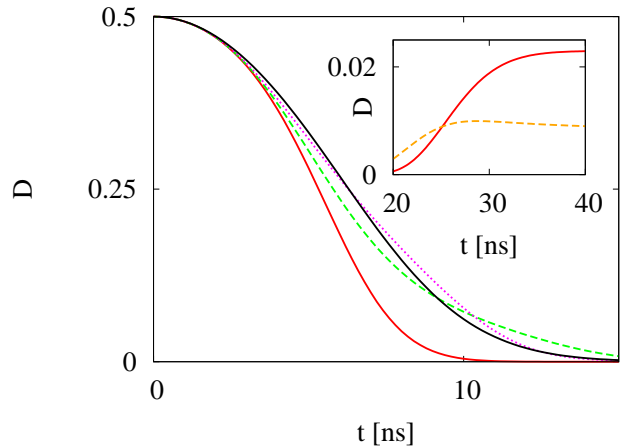}
\caption{\label{Discord}
Time-evolution of the rescaled discord of the initial Bell state
for different magnetic fields: B=0 (solid red line; lower bound on the plots), 
B=11 mT (dashed green line), 
B=16.5 mT (dotted magenta line), B= 1T (blue dashed-dotted line). The inset shows long-time evolution, revealing partial revival of the discord for small magnetic fields: B=0 (solid red line) and B=3 mT (orange dashed line).}
\end{figure}

Fig. (\ref{Discord}) shows the time-evolution of any initial Bell state for different 
magnetic field values. Contrarily to the time-evolution of entanglement 
of the same system \cite{mazurek2013}, oscillations of the rescaled discord $D$
are hardly visible. Furthermore, although the values of $D$
are limited from below by the zero-magnetic-field curve as in the case of entanglement,
they are not limited from above by the infinite-magnetic-field line (contrarily to 
entanglement). This is due to the shape of the decay of the amplitude of the single
coherence present, $|b(t)|$, which is weakly enhanced or slowed by the oscillations of
the QD occupations.  At long time scales,
which are shown in the inset of Fig. (\ref{Discord}), a small revival of the discord
is observed at very low magnetic fields (seen for $B=0$ T and $B=3$ mT), 
which originates from the small revival of the coherence
characteristic of low magnetic fields
and the zero-volume quality of the zero-discord states
which makes discord revivals very common.

It turns out that regardless of the magnetic field, the decay of the discord
for initial Bell states is always confined to the $g(t)\leq 1$ limit, where the
value of the geometric discord is proportional only to the square of the 
amplitude of the coherence (which is then rescaled according to Eq. (\ref{rd})
to get the rescaled discord).

For zero-magnetic-field, although the coherence experiences an involved evolution pattern,
including a revival after the initial strong decay is complete, the coherence and the occupations
always satisfy the relation $g(t)= 1$.
For non-zero magnetic field, the dephasing is faster than the decay of occupations, thus 
$g(t)< 1$ for all times except $t=0$.

The reason for the fact that, for any initial Bell state, the evolution is in the parameter regime $g(t)\leq 1$, can be understood in terms of energy conservation. For simplicity, let us consider
an initial singlet state, 
$|\Psi^- \rangle=1/\sqrt{2}(|\uparrow\downarrow\rangle -|\downarrow\uparrow\rangle)$
and disregard any coherent oscillations present in the system, meaning that $b(t)=b^*(t)$.
The state resulting form
action of the environment-influenced noise on the initial singlet state, rewritten in the basis of projectors into the singlet
$S_{0}$ and triplet $T_{-1}$, $T_{0}$, $T_{1}$ eigenstates of the total angular momentum operator of the two electron system
(where the subscripts correspond to angular momenta projection into $z$ -axis),
is at any given time of the form
\begin{eqnarray}\label{dis}
\nonumber
\rho_{DQD}(t)&=&\left[\frac{1}{2}-a(t)\right] T_{-1}+\left[\frac{1}{2}-a(t)\right] T_{+1} \\
\nonumber
&&+ \left[a(t)+b(t)\right] T_{0} + \left[a(t)-b(t)\right] S_{0}.
\end{eqnarray}
In the zero-magnetic-field regime, there is no mechanism favouring the decay into any of the
triplet subspaces and the evolved state fulfils $\frac{1}{2}-a(t)=a(t)+b(t)$
which is equivalent to the condition $g(t)=1$. Thus the state is always of Werner form, 
\begin{equation}
\nonumber
\rho_{DQD}(t)=[\frac{1}{2}-a(t)]\mathbb{I}+ [4a(t)-1] S_{0}.
\end{equation}
Indeed, Fig. (\ref{obserwabla}) shows constant behaviour of $g(t)=1$ for $B=0$. 
On the other hand, high magnetic field values forbid the process of electron flipping, as there is no first order mechanism to diffuse the energy $2g\mu_B \hat{S}_i^z B$ stemming from the electron flip. Thus the preferable decay channel does not change the angular momentum of the state, which is described by the fact that $|b(t)|$ decays more rapidly than $a(t)$, and the Werner state degeneracy over triplet states projectors is lifted.  Fig. (\ref{obserwabla}) presents this fact by showing the time dependence of $g(t)$ for different magnetic field values. 
Hence, we have shown that for the singlet initial state (this can be generalized
for any initial Bell state; the unitary oscillations do not disturb the discord and do not need to be taken into account) achieving in-differentiable evolution of the rescaled discord is
impossible, since the entrance into the $g(t)\geq 1$ regime, which would lead to it, 
is forbidden by the energy conservation law.

\subsection{Small magnetic field measurement}

\begin{figure}
\includegraphics[scale=0.7]{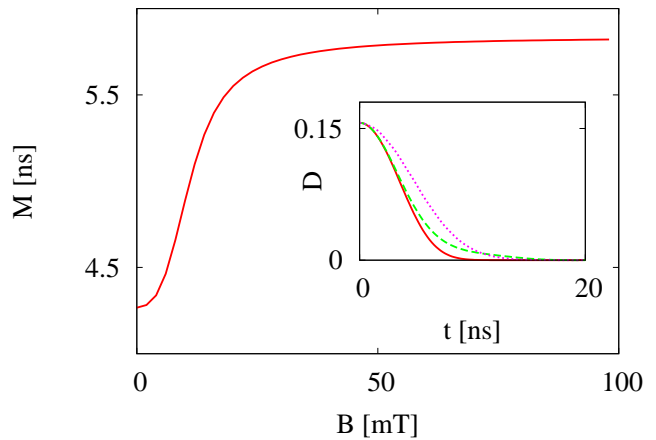}
\caption{\label{separowalne}
Presence of quantum correlations in first $20$ ns of the evolution of the initial separable Werner state
as a function of the magnetic field, quantified by the integral $M(B)$ of Eq. (\ref{M}).
Inset shows discord evolutions for B=0 (red solid line), B=10 mT (green dashed line), B=100 mT (magenta dotted line).
}
\end{figure}

Let us note, that the behaviour of the discord evolution for long times shows strong dependence on small magnetic fields in the range $0-5$ mT (see inset of Fig. \ref{Discord}). 
The ability to perform precise rescaled discord measurements would enable one to detect small magnetic fields; this would not require a precise choice of time of the measurement as long as it is long enough. On the other hand one should not wait too long, since the correlations are expected to decay according to 
$1/\ln t$ \cite{zhang2006,chen2007}. The proposed procedure would expand the region of applicability of a QD magnetic sensor to the region of magnetic fields inaccessible to the entanglement based procedure introduced in Ref.~[\onlinecite{mazurek2013}]. In contrast to the
entanglement based procedure, entanglement is not required as a necessary resource for
these long-time measurements - any initial quantumly correlated Werner state of the form 
\begin{equation}\label{Werner}
\rho_{DQD}(0)=(1-p)\mathbb{I}+ p S_{0}
\end{equation}
would suffice, since its evolution displays asymptotics 
of the same type as the initial singlet state and could be exploited, 
if the measurement precision would be high enough. 
As an example, evolutions of the rescaled discord for a separable state with $p=0.33$ at different magnetic fields are shown in the inset of Fig.\ref{separowalne}.
It is important to note here, that since the evolution always remains in the $g(t)\leq 1$
regime, where it is directly proportional to the amplitude of the single, non-zero coherence
present in the system, measurements of this coherence would suffice
to determine the magnetic field. Hence, the experimental realization of the discord-based
measurement of the magnetic field depends not on the ability to perform direct measurements
of the asymptotic rescaled discord, but on the experimentally attainable precision
of coherence measurements.

\begin{figure}
\includegraphics[scale=0.7]{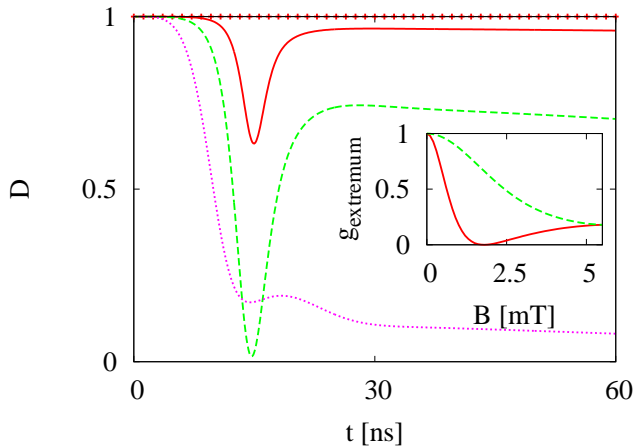}
\caption{\label{obserwabla}
Evolutions of $g(t)$ for B=0 (red points), B=0.5 mT (solid red line), B=1.5 mT (dashed green line) and B=5 mT (dotted magenta line).
Inset shows values of local minimum (solid red line) and local maximum (dashed green line) attained by $g(t)$ during its evolution in different, low magnetic fields.
}
\end{figure}

The idea sketched above is not the only possibility of detecting small magnetic field 
values taking advantage of the rich characteristics of the decoherence driven discord 
evolution. As mentioned previously, Fig. (\ref{obserwabla}) shows the evolution of $g(t)$ for the initial singlet state. Although for zero magnetic field this function is constant, $g(t)=1$, for higher magnetic fields it enters the regime $g(t)<1$ for $t>0$. For small magnetic fields, it
further displays a very strong dependence on the magnetic field. 
It is worth to notice that $g(t)$ can be measured experimentally (although not directly), since 
\begin{equation}
\nonumber
g(t)=\frac{Tr(\sigma_{x}\otimes\sigma_{x}\rho_{DQD}(t))}{Tr(\sigma_{z}\otimes\sigma_{z}\rho_{DQD}(t))},
\end{equation} 
where $\sigma_{i}$, $i=x,y,z$, are the appropriate Pauli matrices.
The shape of $g(t)$ would be a good indicator to qualitatively distinguish between small 
values of the magnetic field, but the depth of the minimum of $g(t)$ could serve as such
an indicator in the range of $0-2$ mT on its own.
Furthermore, from the inset of Fig. (\ref{obserwabla}), which shows the minimum and local maximum
of the function $g(t)$ as a function of the magnetic field, it may be inferred that by registering the value of both parameters for a single magnetic field value,
it would be possible to determine the value of the magnetic field in a wider range of small magnetic fields than using the
entanglement-detection-based scenario. This is because, although the value at the minimum is a non-monotonous
function of the magnetic field, the value at the local maximum decreases monotonously
with the magnetic field
in the studied parameter range. Furthermore, the times at which the minimum and the local maximum
occur are \textit{practically independent} of the magnetic field value at small magnetic fields,
which simplifies the measurement problem.
Again, let us stress that the proposed scheme \textit{does not require entanglement}. 
By investigating initial states of Werner form (\ref{Werner}) it is simple to show 
that the properties of $g(t)$ illustrated in Fig. (\ref{obserwabla}) 
are preserved as long as $p\neq 0$, i.e. are valid also for separable states parametrized by 
$0<p\leq\frac{1}{3}$.  
This is because the amount of noise in the system scales both averages, $Tr(\sigma_{x}\otimes\sigma_{x}\rho_{DQD}(t))$ and $Tr(\sigma_{z}\otimes\sigma_{z}\rho_{DQD}(t))$ 
with the same ratio $p$. 

Let us focus again on the range $B>10$ mT, for which the strong dependence of the time
of entanglement sudden death on the magnetic field enables 
precise sensing of this field. The question is, whether entanglement is necessary
to measure the magnetic field in this regime when utilizing decoherence processes.
To investigate the presence of quantum correlations during the system evolution 
at different magnetic fields, we are clearly unable to follow the idea of 
Ref.~[\onlinecite{mazurek2013}] to measure the dependence of the amount of quantum correlations
present in the system on the magnetic field
by keeping track of changes in the time when there are no quantum correlations left,
since the quantum discord does not experience sudden-death-type behaviour \cite{ferraro2010}. 
Instead, we propose an indicator, 
\begin{equation}
\label{M}
M(B)=\frac{1}{{D}(\rho_{DQD}(0))}\int_{0}^{20 ns} {D}(\rho_{DQD}(t)) dt
\end{equation}
with integration limits arbitrarily chosen in such a way
that the rescaled discord values are reasonably high during the whole 
time under investigation, and that the range of the discord revivals for small B is excluded.
The second requirement is not, however, necessary, and was imposed only in order to obtain a monotonic increase of $M(B)$, which would otherwise attain a local minimum for small
magnetic fields. 
The measure is normalized with respect to  value of $D$ at 
the initial time.
The $M(B)$ dependence on the magnetic field which governs the evolution 
of an initially separable Werner state (\ref{Werner}) with $p=0.33$ is shown in 
Fig. (\ref{separowalne}). It is clear that the magnetic field plays a sustainable role in 
keeping the presence of correlations beyond entanglement in the system during its evolution. Generally speaking, entanglement should not be considered a necessary resource in schemes for sensing $B>10$ mT magnetic fields, similarly as it was concluded for lower magnetic fields. 

\subsection{Indifferentiability of the discord evolution}

\begin{figure}
\includegraphics[scale=0.7]{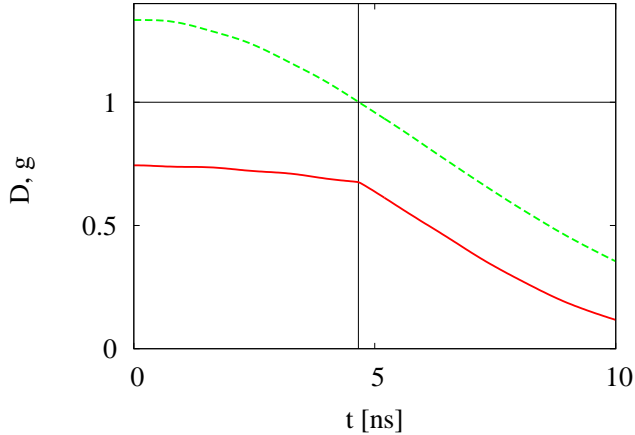}
\caption{\label{nierozniczkowalnosc}
The evolution of $D(\rho_{DQD})$ magnified two times at $B=100$ mT, for a Bell-diagonal 
initial state (\ref{singlet}) with $a=0.4$, $b=0.4$ (red solid line). 
The function is indifferentiable at $t\approx 4.67$ ns, when the value
of $g(t)$ crosses unity (green dashed line).
}
\end{figure}

Finally, let us show that the property of the quantum discord of exhibiting indifferentiable
evolutions \cite{maziero09,mazzola2010} can be displayed by the system of two non-interacting QDs with maximally mixed environments by a proper choice of initial state. Eq. (\ref{dis}) suggests that whenever the equality $g(t)=1$ is attained and crossed through the evolution we should register an indifferentiability of the rescaled discord - indeed, such a situation is attained
for a class of mixed initial states, which keep the form of Eq. (\ref{singlet}),
but for which the occupations are pre-decayed in such a way that 
$g(0)> 1$, meaning that initially the condition $|b(0)|>|1-4a(0)|/2$
must be fulfilled. An example of such an evolution of the 
rescaled discord at $B=100$ mT is shown in Fig. (\ref{nierozniczkowalnosc}), with
$a(0)=0.4$ and $b(0)=0.4$ (red, solid line). Complementarily, the evolution of the
corresponding function $g(t)$ is shown in the same plot (green, dashed line). 
Note, that before the transition, the decay of $D$
is much slower than after the transition and
displays slight oscillations due to the redistribution of the occupations
that occur in the system, which disappear after the transition at $t\approx 4.67$ ns.

\section{Other entangled states \label{sec_non}}

\begin{figure}
\includegraphics[scale=0.7]{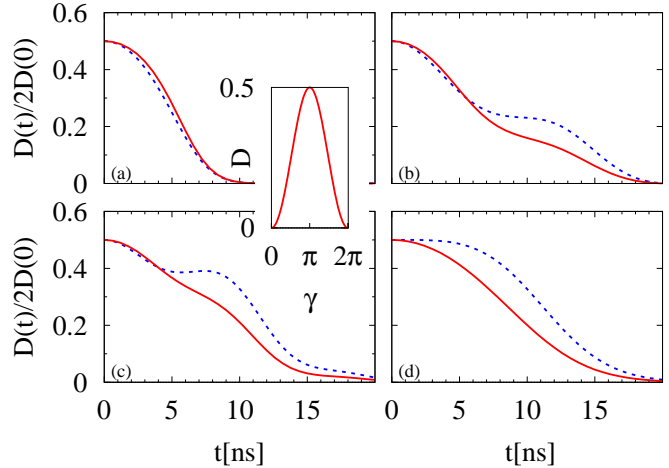}
\caption{\label{nowy1}
The evolution of the rescaled discord
of the initial state of the form (\ref{inif}) normalized to begin at $1/2$ with 
$\gamma=\pi$ (red solid line) and
$\gamma=\pi/2$ (blue short dashed line) for different magnetic fields: $0$ mT (a), $11$ 
mT (b), $16.5$ mT (c), $1$ T (d). 
The inset shows the dependence of rescaled discord of the initial state on the parameter $\gamma$.}
\end{figure}

To complete the study of the evolution of quantum correlations quantified by the
rescaled discord $D$, it is necessary to study non-Bell-diagonal initial states,
for which the full formulas for the upper and lower bounds on the geometric discord,
given by Eqs. (\ref{lower}) and (\ref{upper}), respectively,
need to be used to calculate the lower and upper bounds on $D$ of Eq. (\ref{rd}). 
To gain insight into the possible characteristics of the decay of $D$, it is convenient to analyse the 
behaviour of the initial state of the form
\begin{equation}
\label{inif}
|\psi\rangle =
\frac{1}{2}\left[|00\rangle +|10\rangle 
+|01\rangle +e^{i\gamma}|11\rangle\right].
\end{equation}
This state is, for $e^{i\gamma}=-1$, equivalent to one of the maximally entangled
states given by Eq. (\ref{ini}), with $a=b=1/4$, otherwise it is not a maximally entangled state,
and for $e^{i\gamma}=1$, the state is separable.
For all pure states, the upper and lower bounds on $D$ coincide
and its dependence of the initial state (\ref{inif})
on the phase parameter $\gamma$ is plotted in the inset of Fig. \ref{nowy1}.

\begin{figure}
\includegraphics[scale=0.7]{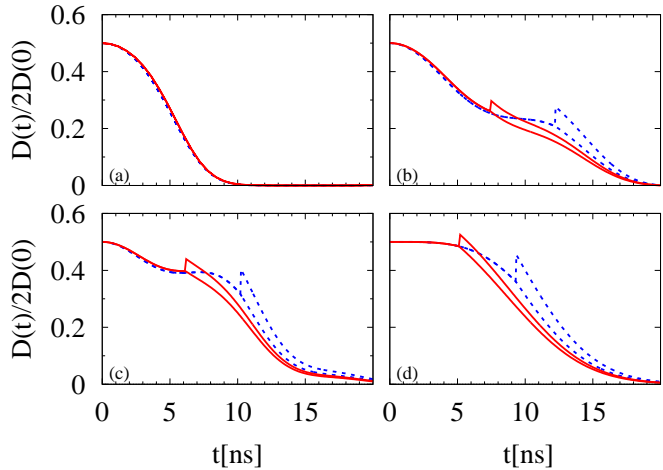}
\caption{\label{nowy2}
The evolution of the lower and upper bounds on the rescaled discord
of the initial state of the form (\ref{inif}) normalized to begin at $1/2$ with 
$\gamma=3\pi/2$ (red solid line) and
$\gamma=7\pi/6$ (blue short dashed line) for different magnetic fields: $0$ mT (a), $11$ 
mT (b), $16.5$ mT (c), $1$ T (d). The lower and upper bounds 
are denoted with the same type of line.}
\end{figure}

Fig. \ref{nowy1} shows the time-evolution of $D$
of the initial state (\ref{inif}) with 
$\gamma=\pi$ (red solid line) and
$\gamma=\pi/2$ (blue short dashed line) for four different values of the magnetic field. 
Analogously, Fig. \ref{nowy2} shows the time-evolution of the rescaled discord
of initial states (\ref{inif}) with $\exp[i\gamma]=(-1+i)/\sqrt{2}$ ($\gamma=3\pi/2$ - red solid line) and
$\exp[i\gamma]=(-1+3i)/\sqrt{2}$ ($\gamma=7\pi/6$ - blue short dashed line).
The evolutions are, for clarity, renormalized in such a way that both discord evolutions start at $1/2$.
In the case of these states, discord oscillations in time are much more pronounced
than in the case of initial Bell states (or even Werner states).
Although the minima and maxima of these oscillations are unaffected by phase factor $\gamma$,
their amplitude is.
Furthermore, it is clearly seen that the evolutions strongly vary depending 
on the phase factors, 
except for the zero magnetic field case when the differences are minuscule.
This is in direct contrast with the evolution of quantum correlations quantified 
by entanglement, for which the same normalization would yield exactly overlapping curves.

Throughout their evolution the initial states corresponding to the plots
of Fig. \ref{nowy1} always display matching lower and upper bounds on the discord.
This is not the case for the initial states corresponding to the plots
of Fig. \ref{nowy2}, for which the lower and upper bounds on the rescaled discord
show a discrepancy after some level of decoherence is reached.
The discrepancy appears earlier at higher magnetic field.
In accordance to the findings of Ref.~[\onlinecite{roszak13}], this occurs in an abrupt
manner, and is followed by a slow decay of the difference between the lower and upper bounds.
These, indifferentiable points in the evolution of the discord upper bound are
also indifferentiable points in the evolution of the discord lower bound
(the jump in the upper bound is accompanied by a transition between two decay curves
in the lower bound.

\section{Conclusion \label{sec_conc}}
We have investigated the evolution of quantum correlations, quantified by the rescaled discord, 
in a system of two electrostatic non-interacting QDs. 
We have shown that for initial Bell states, the system cannot exhibit indifferentiable 
points in the evolution, which are often seen in discord evolutions, due to energy conservation,
regardless of the applied magnetic field.
Such points are visible for initially partially mixed Bell-diagonal states,
and for other classes of pure entangled initial states.
Furthermore, for pure entangled initial states apart from the Bell states,
we have observed a strong dependence of the discord evolutions
on the inter-qubit phase coherence. For imaginary phases, the true value of the 
rescaled discord cannot be found, and only the upper and lower bounds
are available.

We have further studied the magnetic field dependence of the evolutions
in the context of the usability of the double QD system undergoing decoherence due
to an interaction with nuclear spin environments for the measurement of the applied magnetic field.
For initial Bell states, we have found that the sensitivity of the discord
to the magnetic field is of wider range than the sensitivity of entanglement
in the same system. Firstly, the discord displays strong sensitivity to very low magnetic fields
(in the range of $0-5$ mT) for which the sensitivity of entanglement decay is negligible.
To this end we have also pointed out another quantity related to the discord,
which shows strong dependence on very small magnetic fields.
We have also shown that the rescaled discord provides good insight into higher magnetic field values
(of over $10$ mT) for which entanglement sensitivity is optimal.

Lastly, we have shown that regardless of the magnetic field regime,
entanglement is not a necessary resource
for strong magnetic field sensitivity.
Quantum correlations beyond entanglement can serve as a resource for magnetic field sensing in higher magnetic fields, but the range of applicability of a QD sensor can be extended to low magnetic fields.
This is done either by exploiting the properties of the evolution of the correlations present in the system on longer time scales and depending on the magnetic field, or by measurement of the extrema
of the 
observable $g(t)$ which occur at short time scales.
In both regimes, and for all three methods of measuring the magnetic field,
the results can in principle be obtained using separable Werner states
with non-zero discord. 

\begin{acknowledgments}
The authors acknowledge support from the
National Science Centre project 2011/01/B/ST2/05459. 
This work was supported by the TEAM programme of the Foundation for Polish Science co-financed from the European Regional Development Fund (K. R.). P.M. was supported by the Foundation for Polish Science International PhD Projects Programme co-financed by the EU European Regional Development Fund. 
\end{acknowledgments}


\begin{thebibliography}{10}

\bibitem{horodecki2009}
R. {Horodecki}, P. {Horodecki}, M. {Horodecki}, and K. {Horodecki}, Rev. Mod.
  Phys. {\bf 81},  865  (2009).

\bibitem{ollivier2001}
H. {Ollivier} and W.~H. {Zurek}, Physical Review Letters {\bf 88},  017901
  (2002).

\bibitem{henderson2001}
L. {Henderson} and V. {Vedral}, Journal of Physics A Mathematical General {\bf
  34},  6899  (2001).

\bibitem{oppenheim02}
J. {Oppenheim}, M. {Horodecki}, P. {Horodecki}, and R. {Horodecki}, Physical
  Review Letters {\bf 89},  180402  (2002).

\bibitem{modi12}
K. {Modi} {\it et~al.}, Reviews of Modern Physics {\bf 84},  1655  (2012).

\bibitem{knill98}
E. Knill and R. Laflamme, Phys. Rev. Lett. {\bf 81},  5672  (1998).

\bibitem{datta08}
A. Datta, A. Shaji, and C.~M. Caves, Phys. Rev. Lett. {\bf 100},  050502
  (2008).

\bibitem{biham04}
E. Biham, G. Brassard, D. Kenigsberg, and T. Mor, Theoretical Computer Science
  {\bf 320},  15   (2004).

\bibitem{gu12}
M. Gu {\it et~al.}, Nature Physics {\bf 8},  671  (2012).

\bibitem{tekst}
The corresponding effect on the ground of remote state preparation
was found in Ref.~[\onlinecite{dakic12}], where the experiment was
reported together with theoretical evidence for the surprising superiority
of quantum correlations of separable states over entanglement.
However comprehensive approach \cite{horodecki13} has shown its remarkable limitations,
yet leaving space for the expected superiority in cases of restricted
observer's decoding abilities.

\bibitem{dakic12}
B. Dakic {\it et~al.}, Nature Physics {\bf 8},  666  (2012).

\bibitem{horodecki13}
P. Horodecki, J. Tuziemski, P. Mazurek, and R. Horodecki, arXiv:1306.4938, Phys. Rev. Lett.
to be published (2014).

\bibitem{zhang13}
F.-L. Zhang, J.-L. Chen, L.~C. Kwek, and V. Vedral, Scientific Reports {\bf 3},
   2134  (2013).

\bibitem{ferraro2010}
A. Ferraro {\it et~al.}, Phys. Rev. A {\bf 81},  052318  (2010).

\bibitem{dakic2010}
B. {Daki{\'c}}, V. {Vedral}, and {\v C}. {Brukner}, Physical Review Letters
  {\bf 105},  190502  (2010).

\bibitem{luo10}
S. Luo and S. Fu, Phys. Rev. A {\bf 82},  034302  (2010).

\bibitem{nakano13}
T. Nakano, M. Piani, and G. Adesso, Phys. Rev. A {\bf 88},  012117  (2013).

\bibitem{paula13}
F.~M. Paula, T.~R. de~Oliveira, and M.~S. Sarandy, Phys. Rev. A {\bf 87},
  064101  (2013).

\bibitem{spehner13}
D. Spehner and M. Orszag, New Journal of Physics {\bf 15},  103001  (2013).

\bibitem{spehner14}
D. Spehner and M. Orszag, J. Phys. A: Math. Theor. {\bf 47},  035302  (2014).

\bibitem{miranowicz2012}
A. {Miranowicz} {\it et~al.}, \pra {\bf 86},  042123  (2012).

\bibitem{tufarelli13}
T. Tufarelli {\it et~al.}, J. Phys. A {\bf 46},  275308  (2013).

\bibitem{piani12}
M. {Piani}, \pra {\bf 86},  034101  (2012).

\bibitem{elzerman04}
J.~M. Elzerman {\it et~al.}, Nature {\bf 430},  431  (2004).

\bibitem{hanson05}
R. Hanson {\it et~al.}, Physical Rewiev Letters {\bf 94},  196802  (2005).

\bibitem{petta05}
J.~R. Petta {\it et~al.}, Science {\bf 309},  2180  (2005).

\bibitem{laird2010}
E.~A. {Laird} {\it et~al.}, \prb {\bf 82},  075403  (2010).

\bibitem{lai2011}
N.~S. {Lai} {\it et~al.}, Sci. Rep. {\bf 1},  110  (2011).

\bibitem{barthel12}
C. Barthel {\it et~al.}, Phys. Rev. B {\bf 85},  035306  (2012).

\bibitem{medford12}
J. Medford {\it et~al.}, Phys. Rev. Lett. {\bf 108},  086802  (2012).

\bibitem{shulman2012}
M.~D. {Shulman} {\it et~al.}, Science {\bf 336},  202  (2012).

\bibitem{coish2009}
W.~A. {Coish} and J. {Baugh}, Phys. Stat. Sol. B {\bf 246},  2203  (2009).

\bibitem{coish2010}
W.~A. {Coish}, J. {Fischer}, and D. {Loss}, \prb {\bf 81},  165315  (2010).

\bibitem{cywinski2011}
{\L}. {Cywi{\'n}ski}, Acta Phys. Pol. A {\bf 119},  576  (2011).

\bibitem{budker}
D. Budker and M. Romalis, Nat. Physics {\bf 3},  227  (2007).

\bibitem{wasilewski}
W. Wasilewski {\it et~al.}, Phys. Rev. Lett. {\bf 104},  133601  (2010).

\bibitem{maze}
J.~R. Maze {\it et~al.}, Nature {\bf 455},  644  (2008).

\bibitem{mazurek2013}
P. {Mazurek}, K. {Roszak}, R.~W. {Chhajlany}, and P. {Horodecki}, arXiv:1304.1749
(unpublished).

\bibitem{maziero09}
J. Maziero, L.~C. C\'eleri, R.~M. Serra, and V. Vedral, Phys. Rev. A {\bf 80},
  044102  (2009).

\bibitem{mazzola2010}
L. {Mazzola}, J. {Piilo}, and S. {Maniscalco}, Physical Review Letters {\bf
  104},  200401  (2010).

\bibitem{barnes2012}
E. {Barnes}, {\L}. {Cywi{\'n}ski}, and S. {Das Sarma}, Physical Review Letters
  {\bf 109},  140403  (2012).

\bibitem{abragam1983}
A. {Abragam}, {\em {The Principles of Nuclear Magnetism}} (Oxford University
  Press, New York, 1983).

\bibitem{merkulov2002}
I.~A. {Merkulov}, A.~L. {Efros}, and M. {Rosen}, \prb {\bf 65},  205309
  (2002).

\bibitem{barnes2011}
E. {Barnes}, {\L}. {Cywi{\'n}ski}, and S. {Das Sarma}, \prb {\bf 84},  155315
  (2011).

\bibitem{cywinski2009}
{\L}. {Cywi{\'n}ski}, W.~M. {Witzel}, and S. {Das Sarma}, \prb {\bf 79},
  245314  (2009).

\bibitem{liu2007}
R.-B. Liu, W. Yao, and L.~J. Sham, New Journal of Physics {\bf 9},  226
  (2007).

\bibitem{tiersch13}
M. {Tiersch}, G.~G. {Guerreschi}, J. {Clausen}, and H.~J. {Briegel}, 
J. Phys. Chem. A {\bf 118}, 13-20 (2014).

\bibitem{roszak13}
K. Roszak, P. Mazurek, and P. Horodecki, Phys. Rev. A {\bf 87},  062308
  (2013).

\bibitem{blanchard01}
P. Blanchard, L. Jakóbczyk, and R. Olkiewicz, Physics Letters A {\bf 280},  7
   (2001).

\bibitem{horodecki96b}
R. {Horodecki} and M. {Horodecki}, \pra {\bf 54},  1838  (1996).

\bibitem{zhang2006}
W. Zhang {\it et~al.}, Phys. Rev. B {\bf 74},  205313  (2006).

\bibitem{chen2007}
G. {Chen}, D.~L. {Bergman}, and L. {Balents}, \prb {\bf 76},  045312  (2007).

\end{thebibliography}
\end{document}